# Bistability in spatiotemporal mode-locking with dynamic multimode gain

ZHIJIN XIONG,[1,3] YUANKAI GUO,[1,3] WEI LIN,[1,3] HAO XIU,[1] YUNCONG MA,[1] XUEWEN CHEN,[1] ZHAOHENG LIANG,[1] LIN LING,[1] TAO LIU,[1] XIAOMING WEI,[1*] AND ZHONGMIN YANG[1,2*]

[1]*School of Physics and Optoelectronics, State Key Laboratory of Luminescent Materials and Devices, Guangdong Engineering Technology Research and Development Center of Special Optical Fiber Materials and Devices, Guangdong Provincial Key Laboratory of Fiber Laser Materials and Applied Techniques, South China University of Technology, Guangzhou 510640, China*
[2]*Research Institute of Future Technology, South China Normal University, Guangzhou 510006, China*
[3]*These authors equally contributed to this work.*
*email:xmwei@scut.edu.cn; yangzm@scut.edu.cn

**Abstract:** Three-dimensional (3D) dissipative soliton existed in spatiotemporal mode-locked (STML) multimode fiber laser has been demonstrated to be a promising formalism for generating high-energy femtosecond pulses, which unfortunately exhibit diverse spatiotemporal dynamics that have not been fully understood. Completely modeling the STML multimode fiber lasers can shed new light on the underlying physics of the spatiotemporal dynamics and thus better manipulate the generation of high-quality energic femtosecond pulses, which however is still largely unmet. To this end, here we theoretically investigate a dynamic multimode gain model of the STML multimode fiber laser by exploring the multimode rate equation (MMRE) in the framework of generalized multimode nonlinear Schrödinger equation. Using this dynamic multimode gain model, the attractor dissection theory is revisited to understand the dominant effects that determine the modal composition of 3D dissipative soliton. Specifically, by varying the numerical aperture of the multimode gain fiber (MMGF), different gain dynamics that correspond to distinct types of gain attractors are observed. As a result, two distinguishing STML operation regimes, respectively governed by the multimode gain effect and spatiotemporal saturable absorption, are identified. In the latter regime, especially, 3D dissipative solitons present bistability that there exist bifurcated solutions with two different linearly polarized (LP) mode compositions. To verify the theoretical findings, the experimental implementation shows that the state of STML can be switched between different LP modes, and confirms the presence of bistability. Particularly, the 3D-soliton shaping mechanism that is governed by the multimode gain effect is testified for the first time, to the best of our knowledge. It is anticipated that the proposed dynamic multimode gain model can create new potential for studying the complexity of 3D dissipative soliton.

## 1. Introduction

Mode-locked fiber lasers have been intensively investigated in the past decades for their high beam quality, reliability, excellent thermal management, compact design, ease of use, etc [1,2]. Different from the traditional single-mode counterparts, the study of mode-locked multimode fiber lasers is challenging due to the complicated interplay between the transverse and longitudinal modes [3]. Ever since the first demonstration of spatiotemporal mode-locked (STML) in the multimode fiber laser [4], unprecedentedly increasing attention has been drawn as it exhibits to be a promising scheme towards generating high-energy femtosecond pulses [5,6]. The three-dimensional (3D) soliton inherited from the multimode fibers can provide additional degrees of freedom that are promising for various applications [7,8], e.g., nonlinear frequency conversion [9] and spatial multiplexing for dual-comb generation [10]. From the perspective of fundamentals of physics, 3D soliton is one of the 3D optical analogues that can be underpinned by the generic Gross-Pitaevskii equation [11], and shows the potential for interdisciplinary studies of Bose-Einstein condensates [12,13]. While, its extended concept – 3D dissipative soliton that intrinsically prevails in STML multimode fiber laser enables to access the rich 3D nonlinear optical physics [14], like soliton molecules [15,16], self-similar pulses [17], optical rogue waves [18], Q-switching dynamics [19], period-doubling bifurcations [20], spatiotemporal instability [21] and spatial beam cleaning [22,23].

Although many experimental studies of the STML dynamics have been demonstrated, the theoretical understanding of their physics is still limited [24-30]. To this end, L. Wright et al. recently proposed an attractor dissection theory for dissecting the intracavity effects responsible for forming 3D dissipative soliton [31,32]. Soon after, the STML laser using a step-index multimode fiber that has a large modal dispersion [33] was implemented, and the underlying mechanism was discussed afterwards [34], in which the interaction between the spatiotemporal saturable absorption (SA) and spatial gain competition was analyzed. It is noticed that the theoretical modeling by far mostly hinges on the generalized multimode nonlinear Schrödinger equation (GMMNLSE), wherein a Gaussian gain shaping is adopted at each spatial point of the mode field, i.e., termed as Gaussian-gain model [4,7,14,31,33,34]. Despite the simplicity, it fails to include several ingredients essential for these spatiotemporal dynamics that are commonly observed in experiments, such as pump depletion, spatial gain saturation, and explicit gain filtering [35]. To this end, a supplementary model for STML lasers that leverages the multimode rate equation (MMRE) is highly desired [36-38], as its feasibility has been verified by the unidirectional propagation of 3D solitons in a multimode fiber amplifier, which involves spatiotemporal deterioration [39] and beam self-cleaning [35].

In this work, we present a comprehensive model for studying the intracavity nonlinear dynamics of STML multimode fiber lasers by exploring the dynamic multimode gain in a framework of integrating the mode-resolved MMRE with GMMNLSE. Using this new model of the STML laser, the attractor dissection theory is revisited to investigate the modal dynamics of the 3D dissipative soliton. Specifically, as the 3D dissipative soliton circulates in the STML cavity modeled with dynamic multimode gain, different gain dynamics can yield distinct types of gain attractors, resulting in two operation regimes of STML that are dominated by the multimode gain effect and spatiotemporal SA, respectively. In the latter regime, particularly, the 3D dissipative soliton presents bistability — bifurcated solutions with two different linearly polarized (LP) mode compositions, i.e., $LP_{01}$ mode and $LP_{11}$ mode, respectively. In the experiment, we further confirmed the switching between $LP_{01}$ and $LP_{11}$ modes. The proposed dynamic multimode gain model of STML multimode fiber lasers might open new opportunities for fully studying the spatiotemporal dynamics of 3D dissipative solitons.

## 2. Dynamic multimode gain model of STML multimode fiber laser

Figure 1 conceptually illustrates the modeling of the STML multimode fiber laser that leverages dynamic multimode gain. As shown in Fig. 1(a), the STML cavity mainly consists of a multimode gain fiber (MMGF), a SA and an optical coupler (OC). The MMGF is modeled as a quasi-three-level system in the framework of GMMNLSE with mode-resolved MMRE, which can fully interpret the multimode gain dynamics interplayed with nonlinear modal coupling. Here, the MMGF is a step-index multimode $Yb^{3+}$-doped fiber (YDF) that has a core diameter of 20 μm. In the modeling, six transverse modes (TMs) are considered, including $LP_{01}$, $LP_{11a}$, $LP_{11b}$, $LP_{21a}$, $LP_{21b}$, and $LP_{02}$. To understand the fundamental mechanism of generating stable 3D dissipative solitons, the attractor dissection theory is associated with this new model to analyze the modal compositions of eigenpulses with respect to the gain operator ($\widehat{G}$) and a joint effect operator ($\widehat{D} + \widehat{SA}$, $\widehat{D}$ and $\widehat{SA}$ account for the dispersion and SA effect, respectively), as depicted in Fig. 1(b). Under the interplay between the $\widehat{G}$ and ($\widehat{D} + \widehat{SA}$) attractors, the formation of 3D dissipative solitons in the laser cavity can be divided into two regimes (i and ii), led by multimode gain effect and spatiotemporal SA, respectively. In regime i, the $LP_{01}$ mode dominates the gain extraction process, resulting in monostability. In regime ii, the multimode gain is homogenized for different spatial modes, and the presence of $(\widehat{D}+\widehat{SA})$ attractors gives rise to the bistability (to be discussed).

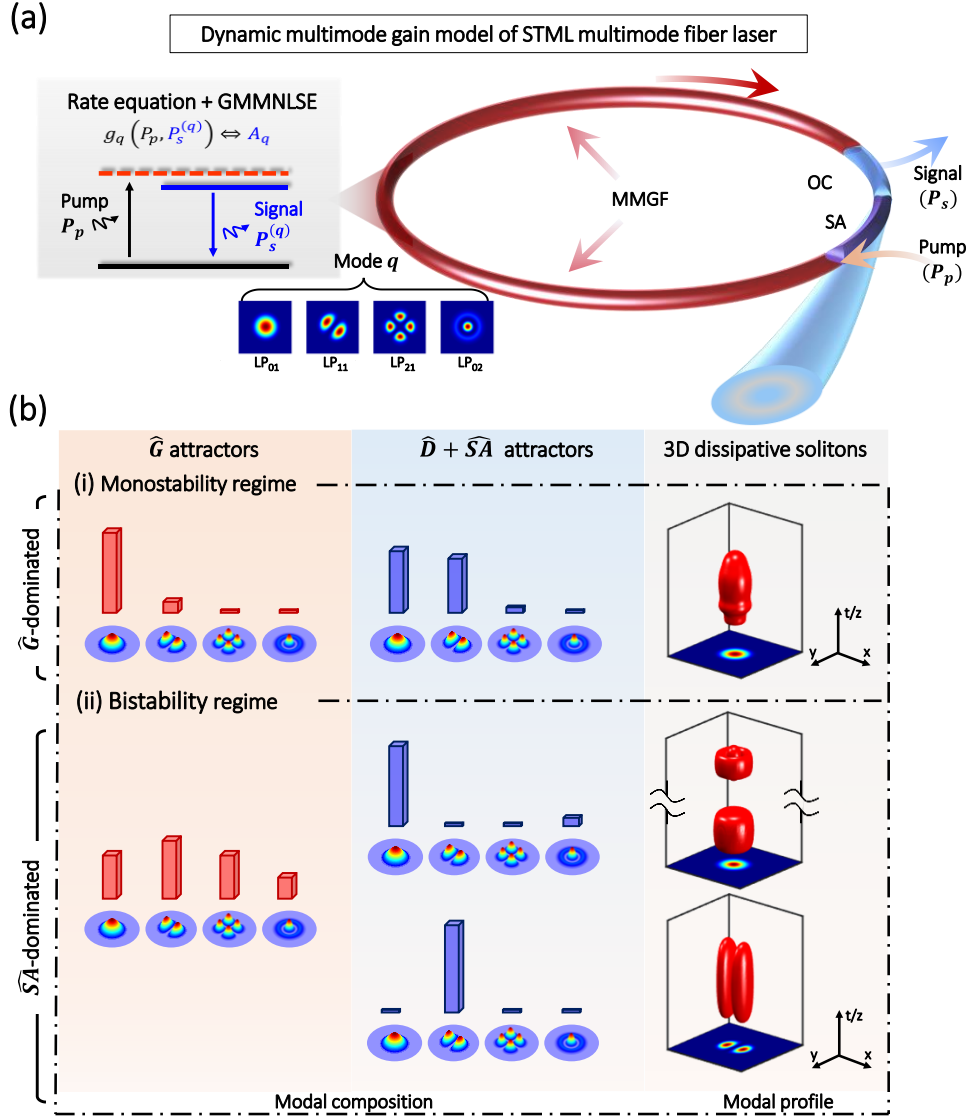

**Fig. 1**. Dynamic multimode gain model of spatiotemporal mode-locked (STML) multimode fiber laser. (a) Schematic diagram of the STML multimode fiber laser. The dynamic multimode gain of the multimode gain fiber (MMGF) is modeled by the generalized multimode nonlinear Schrödinger equation (GMMNLSE) that is associated with the multimode rate equation (MMRE). Here, six transverse modes (TMs, including two pairs of degenerate modes) are considered in the numerical study. OC: optical coupler; SA: saturable absorption. (b) Formation of 3D dissipative solitons in the STML multimode fiber laser. Here, the 3D dissipative solitons are generated by the combined role of $\widehat{G}$ and $(\widehat{D}+\widehat{SA})$, wherein $\widehat{G}$, $\widehat{D}$, and $\widehat{SA}$ represent the operators of gain, dispersion, and SA, respectively. Two regimes, i.e., (i) monostability and (ii) bistability dominated by the multimode gain effect and spatiotemporal SA, respectively. The height of the red and blue bars quantifies the modal compositions of the $\widehat{G}$ and $(\widehat{D}+\widehat{SA})$ attractors, respectively.

**A. Dynamic multimode gain**

In the framework of GMMNLSE characterizing the propagation of 3D dissipative solitons within the multimode fiber laser cavity using a MMGF, the gain function $g_q$ is described by the mode-resolved MMRE, which contrasts with the multimode gain presented in full-field formalism [4,31]. Specifically, a set of multimode propagation equations can be expressed in the forms of

$$\frac{dP_p(z)}{dz} = \Gamma_p \left(\sigma_{ep} n_2(z) - \sigma_{ap} n_1(z)\right) P_p(z), \tag{1}$$

$$\frac{dP_s^{(q)}(z,\lambda_k)}{dz} = \sigma_{es}(\lambda_k)\left[P_s^{(q)}(z,\lambda_k) + 2h\frac{c^2}{\lambda_k^3}\Delta\lambda\right]\Gamma_s^{(q)} n_2(z) - \sigma_{as}(\lambda_k) P_s^{(q)}(z,\lambda_k)\Gamma_s^{(q)} n_1(z), \tag{2}$$

where, $P_p(z)$ and $P_s^{(q)}(z,\lambda_k)$ represent the $z$-dependent pump power and signal power of the spatial mode $q$ at the wavelength of $\lambda_k$. $n_2$ and $n_1$ denote the population densities of Yb$^{3+}$ ions at the upper and lower energy levels, respectively. More details about the dynamic multimode gain model are provided in Sec. S1.1 of Supplemental Material [40], and the definitions and values of key parameters are summarized in Table S1.

From Eq. (2), we arrive at an expression of the mode-resolved gain function $g_q$, i.e.,

$$g_q(\omega_k; z) = \Gamma_s^{(q)}\left(\sigma_{es}(\lambda_k) n_2(z) - \sigma_{as}(\lambda_k) n_1(z)\right). \tag{3}$$

### B. Attractor dissection

To unveil the key intracavity effects responsible for generating the 3D dissipative soliton in the proposed dynamic multimode gain model, the attractor dissection theory is leveraged to evaluate the two principal ingredients, i.e., the mode-resolved gain and SA effect. These two effects are regarded as projection operators that represent the transformations of the optical field envelopes [31], and the corresponding field attractors $A(x,y,t)$ (also understood as 'eigenpulses') with respect to the operators can be expressed as

$$A_q^G(t) = \lim_{n\to\infty}\left[\hat{R}\hat{G}\right]^n E_q^0(t), \tag{4}$$

$$A^{SA}(x,y,t) = \lim_{n\to\infty}\left[\hat{R}\widehat{SA}\hat{D}\right]^n E^0(x,y,t), \tag{5}$$

where, $A^G$ and $A^{SA}$ are the eigensolutions with regard to the MMGF and SA. $\hat{R}$ is a rescaling operator to retain the total energy. $E_q^0(t)$ and $E^0(x,y,t)$ are the initial fields in the mode-resolved and full-field formalisms, respectively. A full description of the attractor dissection theory can be found in the Sec. S1.2 of Supplemental Material [40]. For intuitive understanding, Fig. 1(b) showcases two representative modal compositions of the $\hat{G}$ and $(\hat{D}+\widehat{SA})$ attractors. Regarding the attractor, there are several noteworthy aspects: (1) the use of operator $\hat{G}$ provides an approximate way to describe the propagation of 3D dissipative solitons in the MMGF, although it is too complicated in reality to unveil the interplay between the multimode gain, spatiotemporal dispersion and nonlinear modal coupling (also discussed in Sec. S2 of Supplemental Material [40]); (2) the modal composition of the field attractor is interpreted as the energy distribution of the TM families. In this way, the spatiotemporal dispersion and nonlinear modal coupling less influence the modal dynamics and thus justify the use of operator $\hat{G}$ (see Sec. S2.2 of Supplemental Material for more details [40]); (3) the modal dispersion is crucial to assess the $\widehat{SA}$ attractor and achieve convergence, since the induced temporal walk-off effect may influence the operator $\widehat{SA}$; (4) the field attractor, particularly $\widehat{SA}$, can manifest the multistability and is then, to some extent, susceptible to the initial condition.

## 3. Bistability dynamics of 3D dissipative soliton governed by dynamic multimode gain

### A. Transition from monostability to bistability

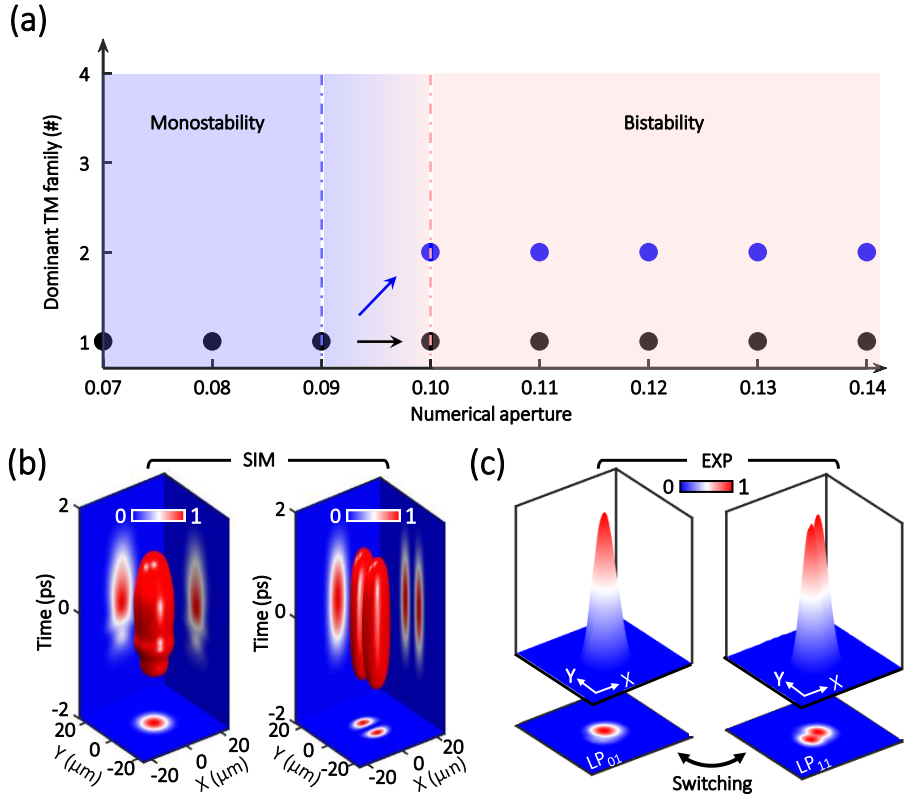

**Fig. 2.** Spatiotemporal dynamics of 3D dissipative soliton in the STML multimode fiber laser with dynamic multimode gain. (a) Dominant spatial modes of the 3D dissipative soliton as the numerical aperture (NA) of the MMGF varies from 0.07 to 0.14. The monostability and bistability regimes are recognized, revealing that 3D dissipative soliton bifurcates from a single attractor (stable node) to pairwise attractors, as it is susceptible to the initial conditions. Here, the TM families 1, 2, 3, and 4 correspond to $LP_{01}$, $LP_{11}$, $LP_{21}$, and $LP_{02}$ modes, respectively. (b) Spatiotemporal profiles of the 3D dissipative solitons for NA = 0.07 (left) and 0.14 (right), respectively. The 3D isosurface plots of the spatiotemporal profiles are set to 10% of the peak intensity. (c) Beam profiles of bistability states in the experiment. The $LP_{01}$ and $LP_{11}$ modes can be experimentally switched between each other according to the initial conditions.

By utilizing the proposed dynamic multimode gain model, we here explore the generation of 3D dissipative solitons in the STML multimode fiber laser. Consistent with the propagation in MMGF (see details in Sec. S2 of Supplemental Material [40]), the numerical aperture (NA) of the MMGF is varied to access distinctive multimode gain characteristics and unveil the underlying mechanism of the spatiotemporal dynamics of generating 3D dissipative solitons. Fig. 2(a) shows that two distinguishing regimes, which are featured by monostability and bistability, respectively, are identified as the NA of the MMGF varies from 0.07 to 0.14, while the corresponding numerical results are provided in Sec. S3 of Supplemental Material [40]. Specifically, when the fundamental mode dominates the gain extraction interpreted by the $\hat{G}$ attractor, the modal dynamics exhibit a monopolistic state dominated by the $LP_{01}$ mode, termed as the monostability regime. When the gain competition between modes is strong, the modal dynamics present two states wherein the $LP_{01}$ or $LP_{11}$ modes prevail over other spatial modes, termed as the bistability regime. Regarding a dissipative 3D system, the modal dynamics of the 3D dissipative soliton in the phase space manifest two domains. The first domain is defined by the initial conditions that asymptotically approach the $LP_{01}$-dominated attractor, and such convergence holds for all NA values in the range of [0.07, 0.14], as denoted by the black arrow in Fig. 2(a). The second domain behaves as a local bifurcation, while the spatial mode $LP_{01}$ interchanges its stability with $LP_{11}$ once the multimode gain exhibits a relatively uniform distribution, as denoted by the blue arrow in Fig. 2(a). The spatiotemporal characteristics of the two attractors are further illustrated in Fig. 2(b). The $LP_{01}$ mode, together with weak $LP_{02}$ mode with a considerably low energy ratio, represents the first attractor, which can also be identified by the 3D isosurface plot shown in the left panel of Fig. 2(b). For the second attractor, the 3D isosurface plot identifies a single-$LP_{11}$ modal composition in the right panel Fig. 2(b), i.e., the degenerate $LP_{11a}$ and $LP_{11b}$ modes with energy ratios of 0.18 and 0.82, respectively. Then, a STML multimode fiber laser is implemented to explore the bistability dynamics, and we observe the switching process between the $LP_{01}$ and $LP_{11}$ modes in the STML state, with their beam profiles shown Figs. 2(c). More experimental details are provided in the Sec. S4 of Supplemental Material [40].

## B. Mechanism based on attractor dissection theory

To gain more insight into the spatiotemporal dynamics of the 3D dissipative solitons with dynamic multimode gain, we focus on two typical cases with NA = 0.07 and 0.14, i.e., corresponding to the monostability and bistability regimes, respectively. In the monostability regime with NA = 0.07, the evolution of the energy distribution of the TM families is shown in Fig. 3(a), wherein it presents a rapid modal convergence towards the $LP_{01}$ mode, which can be verified by a near-Gaussian beam profile at roundtrip (RT) #2, as shown in the inset of Fig. 3(a). It is noticed that the $LP_{01}$ mode sheds a small fraction of energy into the $LP_{02}$ mode through the inter-LP-mode FWM. To understand such modal dynamics, we access the intracavity evolution of the energy distribution in the early stage of the RT evolution, i.e., RTs #2 and #3 in this case. The left panel of Fig. 3(b) illustrates the variation of the mode components (only the $LP_{01}$ and $LP_{11}$ modes are shown here) as the 3D dissipative soliton passes through each cavity element, i.e., MMGF, OC and SA. The results identify that there exist two important processes responsible for the beam shaping: 1) in the MMGF, the fundamental mode $LP_{01}$ has advantages in the energy extraction, and thus prevails in the gain competition by continuously increasing its energy ratio. After passing through the OC and SA, the 3D dissipative soliton experiences a significant energy amplification for cavity-loss compensation, in which the energy ratio is further increased in the second MMGF; 2) the spatiotemporal SA accompanies with the multimode gain dynamics to increase the energy ratio of the $LP_{01}$ mode. As a complementary approach, the attractor dissection theory that is useful for tracing the crucial intracavity effects [31] can be applied. The right panel of Fig. 3(b) shows the acquired field attractors of the operators $\hat{G}$ and $(\hat{D} + \widehat{SA})$, such that the energy distributions of the TM families can be quantified. The calculated $LP_{01}$-dominated attractor suggests that the multimode gain (i.e., the operator $\hat{G}$) plays a decisive role in the spatial shaping, while the SA [i.e., the operator $(\hat{D} + \widehat{SA})$] accelerates this process. In contrast to the modal shaping of the 3D dissipative soliton influenced by the interplay between multimode gain and SA, the temporal shaping of the 3D dissipative soliton that entails intensity discrimination can be merely realized by the SA. Fig. 3(c) illustrates the spatiotemporal profiles (3D isosurface plots) of the 3D dissipative solitons at RTs #3, 5, 10, and 30. The temporal shaping of the 3D dissipative soliton is asynchronous with that of modal shaping (see projections on the *y-t* and *x-t* planes), i.e., lagging behind until RT #30. It is intriguing to recognize this asynchronous spatial and temporal shaping of the 3D dissipative soliton in a multimode fiber laser.

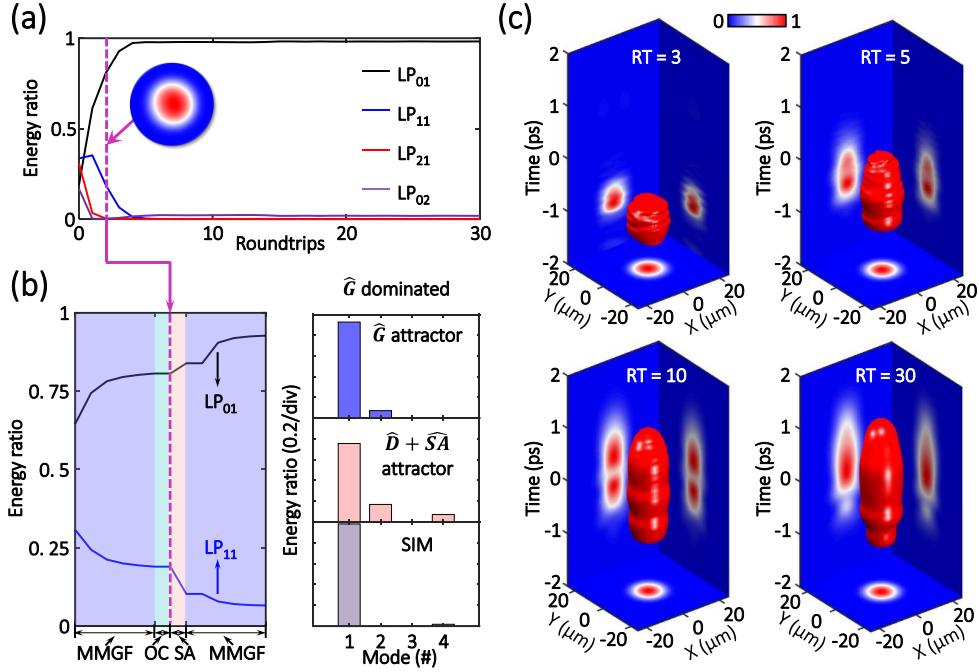

**Fig. 3.** Spatiotemporal dynamics of the 3D dissipative soliton generated from a MMGF with NA = 0.07. (a) Evolution of the energy distribution of the TM families. Inset depicts the beam profile at roundtrip (RT) #2. (b) Intracavity evolution of the energy ratios of the $LP_{01}$ and $LP_{11}$ modes (left panel). Right panel shows that the modal compositions of the 3D dissipative soliton in simulation (labeled by SIM), $\hat{G}$ and $(\hat{D} + \widehat{SA})$ attractors. The results extracted from the OC at RT #2 are indicated by the dashed lines in (a) and (b). The MMGF, OC and SA are the components of the STML multimode fiber laser. Note that, the second MMGF belongs to the next RT, i.e., RT #3. (c) RT-evolving spatiotemporal profiles of the 3D dissipative soliton in the buildup state.

In the bistability regime with NA = 0.14, different initial conditions can probe different steady states of the 3D dissipative solitons. Figs. 4(a) and (b) show the evolutions of the energy distributions. It is worth noting that the energy distributions of the TM families similarly evolve in the beginning, but diverge after RT #4, yielding distinguishing beam profiles, as shown in the insets of Figs. 4(a) and (b). To clarify its origin, the intracavity evolution of the energy ratios of the $LP_{01}$ and $LP_{11}$ modes are shown in the left panels of Figs. 4(c) and (d). In contrast to the case shown in Fig. 3, here the multimode gain plays a trivial role in the modal dynamics of the 3D dissipative soliton, while the spatiotemporal SA mainly leads to the abrupt change of the energy ratios of the $LP_{01}$ and $LP_{11}$ modes. The above results suggest that the SA plays a dominant role in the bistability of the STML laser. From the perspective of physical understanding, it is interesting to dissect the versatility of the SA. To this end, the full-field light fields extracted from RTs #4 and #5 are employed as the initial conditions $E^0(x, y, t)$ for calculating the field attractors of the operator $(\hat{D}+\widehat{SA})$, respectively. In the meantime, the field attractors of the operator $\hat{G}$ are calculated when weak noisy fields are employed as the initial conditions in the dynamic multimode gain model. As showcased in the right panels of Figs. 4(c) and (d), the $\hat{G}$ attractors exhibit similar modal components, while the $(\hat{D}+\widehat{SA})$ attractors present dramatically different energy distributions. The numerical simulations using the dynamic multimode gain model, denoted as SIM in Figs. 4(c) and (d), indicate a good consistency with that of the $(\hat{D}+\widehat{SA})$ attractor, suggesting that the bistability of the 3D dissipative soliton is originated from the multistable nature of the spatiotemporal SA.

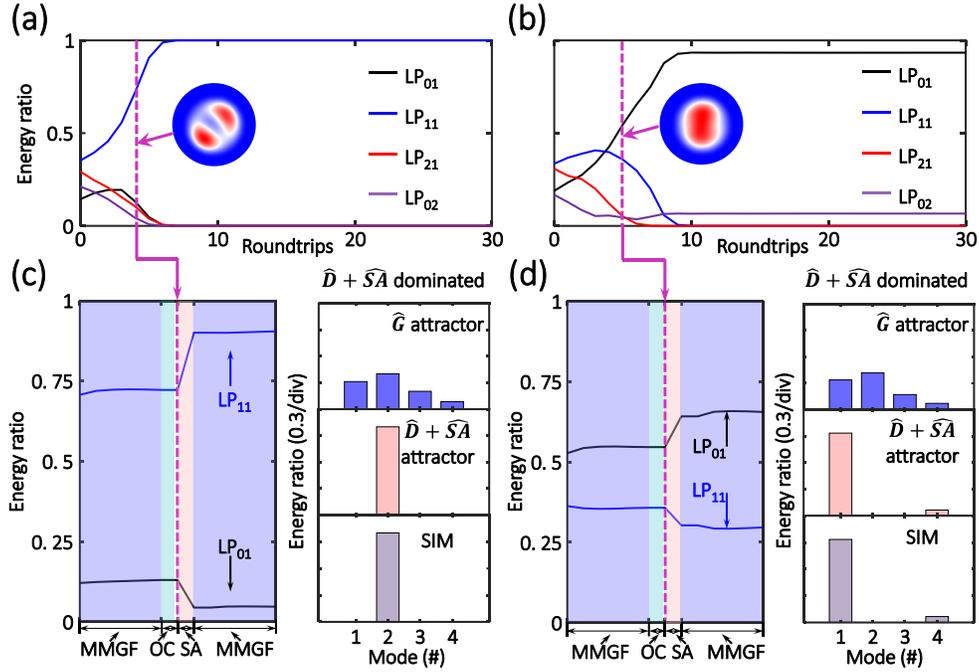

**Fig. 4.** Spatiotemporal dynamics of the 3D dissipative solitons generated from a MMGF with NA = 0.14. (a,b) Evolutions of the energy distributions of the TM families for two different 3D dissipative solitons that are dominated by $LP_{11}$ (a) and $LP_{01}$ (b) modes, respectively. Insets depict the beam profiles at RTs #4 and #5 of the two cases, respectively. (c,d) Intracavity evolutions of the energy ratios of the $LP_{01}$ and $LP_{11}$ modes (left panel). Right panels show that modal compositions of the 3D dissipative solitons in the simulation (labeled by SIM), $\hat{G}$ and $(\hat{D} + \widehat{SA})$ attractors. The results extracted from the OC at RT #4 and #5 are indicated by the dashed lines in (a,c) and (b,d), respectively.

## 4. Conclusion and outlook

In conclusion, we comprehensively explore the spatiotemporal dynamics in STML multimode fiber laser by developing a dynamic multimode gain model that leverages MMRE and GMMNLSE. To unveil the formation mechanism of the 3D dissipative soliton, we revisit the attractor dissection theory to identify the roles of key intracavity ingredients in the shaping of 3D dissipative solitons, including the multimode gain and spatiotemporal SA. Specifically, via varying the NA of the MMGF, diverse gain dynamics, as well as different gain attractors are revealed, giving rise to two operation regimes — governed by multimode gain effect and spatiotemporal SA, respectively. The transition mechanism from monostability to bistability is then theoretically and experimentally investigated, and the bifurcated solutions with two distinct LP mode compositions, i.e., $LP_{01}$ and $LP_{11}$ modes, are identified. The proposed dynamic multimode gain model opens up new possibilities for investigating the spatiotemporal dynamics of 3D dissipative solitons in multimode fiber lasers, and understanding the other 3D dissipative system.

**Funding.** This work was partially supported by National Natural Science Foundation of China (NSFC) (62375087 and 12374304), Key-Area Research and Development Program of Guangdong Province (2023B0909010002), NSFC Development of National Major Scientific Research Instrument (61927816), Introduced Innovative Team Project of Guangdong Pearl River Talents Program (2021ZT09Z109), Natural Science Foundation of Guangdong Province (2021B1515020074), and Science and Technology Project of Guangdong (2020B1212060002).

**Disclosures.** The authors declare no conflict of interest.

**Data availability.** Data underlying the results presented in this paper are not publicly available at this time but may be obtained from the authors upon reasonable request.

## References

1. D. Brida, G. Krauss, A. Sell, and A. Leitenstorfer, "Ultrabroadband Er:fiber lasers," Laser Photonics Rev. **8**(3), 409–428 (2014).
2. J. Kim and Y. Song, "Ultralow-noise mode-locked fiber lasers and frequency combs: principles, status, and applications," Adv. Opt. Photon. **8**(3), 465–540 (2016).
3. D. Côté and H. M. van Driel, "Period doubling of a femtosecond Ti:sapphire laser by total mode locking," Opt. Lett. **23**(9), 715–717 (1998).
4. L. G. Wright, D. N. Christodoulides, and F. W. Wise, "Spatiotemporal mode-locking in multimode fiber lasers," Science **358**(6359), 94–97 (2017).
5. W. Fu, L. G. Wright, P. Sidorenko, S. Backus, and F. W. Wise, "Several new directions for ultrafast fiber lasers [Invited]," Opt. Express **26**(8), 9432–9463 (2018).
6. B. Cao, C. Gao, K. Liu, X. Xiao, C. Yang, and C. Bao, "Spatiotemporal mode-locking and dissipative solitons in multimode fiber lasers," Light Sci. Appl. **12**, 260 (2023).
7. J. C. Jing, X. Wei, and L. V. Wang, "Spatio-temporal-spectral imaging of non-repeatable dissipative soliton dynamics," Nat. Commun. **11**, 2059 (2020).
8. Y. Guo, X. Wen, W. Lin, W. Wang, X. Wei, and Z. Yang, "Real-time multispeckle spectral-temporal measurement unveils the complexity of spatiotemporal solitons," Nat. Commun. **12**, 67 (2021).
9. L. Wright, D. Christodoulides, and F. Wise, "Controllable spatiotemporal nonlinear effects in multimode fibers," Nat. Photon. **9**, 306–310 (2015).
10. E. Lucas, G. Lihachev, R. Bouchand, N. Pavlov, A. Raja, M. Karpov, M. Gorodetsky, and T. J. Kippenberg, "Spatial multiplexing of soliton microcombs," Nat. Photon. **12**, 699–705 (2018).
11. A. Fusaro, J. Garnier, K. Krupa, G. Millot, and A. Picozzi, "Dramatic acceleration of wave condensation mediated by disorder in multimode fibers," Phys. Rev. Lett. **122**(12), 123902 (2019).
12. K. Krupa, A. Tonello, B. M. Shalaby, M. Fabert, A. Barthélémy, G. Millot, S. Wabnitz and V. Couderc, "Spatial beam self-cleaning in multimode fibres," Nat. Photonics **11**, 237–241 (2017).
13. H. Pourbeyram, P. Sidorenko, F. O. Wu, N. Bender, L. Wright, D. N. Christodoulides, and F. Wise, "Direct observations of thermalization to a Rayleigh–Jeans distribution in multimode optical fibres," Nat. Phys. **18**(6), 685–690 (2022).
14. L. G. Wright, F. O. Wu, D. N. Christodoulides, and F. W. Wise, "Physics of highly multimode nonlinear optical systems," Nat. Phys. **18**(9), 1018–1030 (2022).
15. Y. Guo, W. Lin, W. Wang, R. Zhang, T. Liu, Y. Xu, X. Wei, and Z. Yang, "Unveiling the complexity of spatiotemporal soliton molecules in real time," Nat. Commun. **14**, 2029 (2023).
16. K. Liu, X. Xiao, Y. Ding, H. Peng, D. Lv, and C. Yang, "Buildup dynamics of multiple solitons in spatiotemporal mode-locked fiber lasers," Photon. Res. **9**(10), 1898–1906 (2021).
17. U. Teğin, E. Kakkava, B. Rahmani, D. Psaltis, and C. Moser, "Spatiotemporal self-similar fiber laser," Optica **6**(11), 1412–1415 (2019).
18. U. Teğin, P. Wang, and L. V. Wang, "Real-time observation of optical rogue waves in spatiotemporally mode-locked fiber lasers," Commun. Phys. **6**, 60 (2023).
19. K. Liu, X. Xiao, and C. Yang, "Observation of transition between multimode Q-switching and spatiotemporal mode locking," Photon. Res. **9**(4), 530–534 (2021).
20. X. Xiao, Y. Ding, S. Fan, X. Zhang, and C. Yang, "Spatiotemporal period-doubling bifurcation in mode-locked multimode fiber lasers," ACS Photonics **9**(12), 3974–3980 (2022).
21. Y. Sun, P. Rivas, C. Milián, Y. Kartashov, M. Ferraro, F. Mangini, R. Jauberteau, F. Talenti, and S. Wabnitz, "Robust three-dimensional high-order solitons and breathers in driven dissipative systems: a Kerr cavity realization," Phys. Rev. Lett. **131**, 137201 (2023).
22. U. Tegin, B. Rahmani, E. Kakkava, D. Psaltis, and C. Moser, "Single-mode output by controlling the spatiotemporal nonlinearities in mode-locked femtosecond multimode fiber lasers," Adv. Photonics **2**(5), 056005 (2020).
23. G. Fu, T. Qi, W. Yu, L. Wang, Y. Wu, X. Pan, Q. Xiao, D. Li, M. Gong, and P. Yan, "Beam self-cleaning of 1.5 μm high peak-power spatiotemporal mode-locked lasers enabled by nonlinear compression and disorder," Laser Photon. Rev. **17**(7), 2200987 (2023).
24. J. Chen, W. Hong, and A. Luo, "Nonlinear dynamics of beam self-cleaning on $LP_{11}$ mode in multimode fibers," Opt. Express **30**(24), 43453–43463 (2022).
25. X. Wei, J. C. Jing, Y. Shen, and L. V. Wang, "Harnessing a multi-dimensional fibre laser using genetic wavefront shaping," Light Sci. Appl. **9**, 149 (2020).
26. T. Mayteevarunyoo, B. A. Malomed, and D. V. Skryabin, "Spatiotemporal dissipative solitons and vortices in a multi-transverse-mode fiber laser," Opt. Express **27**(26), 37364–37373 (2019).
27. V. L. Kalashnikov and S. Wabnitz, "Distributed Kerr-lens mode locking based on spatiotemporal dissipative solitons in multimode fiber lasers," Phys. Rev. A **102**(2), 023508 (2020).
28. H. Zhang, Y. Zhang, J. Peng, X. Su, X. Xiao, D. Xu, J. Chen, T. Sun, K. Zheng, J. Yao, and Y. Zheng, "All-fiber spatiotemporal mode-locking lasers with large modal dispersion," Photon. Res. **10**(2), 483–490 (2022).
29. B. Cao, C. Gao, Y. Ding, X. Xiao, C. Yang, and C. Bao, "Self-starting spatiotemporal mode-locking using Mamyshev regenerators," Opt. Lett. **47**(17), 4584–4587 (2022).

30. M. Zitelli, V. Couderc, M. Ferraro, F. Mangini, P. Parra-Rivas, Y. Sun, and S. Wabnitz, "Spatiotemporal mode decomposition of ultrashort pulses in linear and nonlinear graded-index multimode fibers," Photon. Res. **11**(5), 750–756 (2023).
31. L. G. Wright, P. Sidorenko, H. Pourbeyram, Z. M. Ziegler, A. Isichenko, B. A. Malomed, C. R. Menyuk, D. N. Christodoulides, and F. W. Wise, "Mechanisms of spatiotemporal mode-locking," Nat. Phys. **16**, 565–570 (2020).
32. F. Ö. Ilday, "Mode-locking dissected." Nat. Phys. **16**(5), 504–505 (2020).
33. Y. Ding, X. Xiao, K. Liu, S. Fan, X. Zhang, and C. Yang, "Spatiotemporal mode-locking in lasers with large modal dispersion," Phys. Rev. Lett. **126**, 093901 (2021).
34. C. Gao, B. Cao, Y. Ding, X. Xiao, D. Yang, H. Fei, C. Yang, and C. Bao, "All-step-index-fiber spatiotemporally mode-locked laser," Optica **10**(3), 356–363 (2023).
35. Y. H. Chen, H. Haig, Y. Wu, Z. Ziegler, and F. Wise, "Accurate modeling of ultrafast nonlinear pulse propagation in multimode gain fiber," J. Opt. Soc. Am. B **40**(10), 2633–2642 (2023).
36. M. Gong, Y. Yuan, C. Li, P. Yan, H. Zhang, and S. Liao, "Numerical modeling of transverse mode competition in strongly pumped multimode fiber lasers and amplifiers," Opt. Express **15**(6), 3236–3246 (2007).
37. Z. Jiang and J. R. Marciante, "Impact of transverse spatial-hole burning on beam quality in large-mode-area Yb-doped fibers," J. Opt. Soc. Am. B **25**(2), 247–254 (2008).
38. L. Huang, L. Kong, J. Leng, P. Zhou, S. Guo, and X. Cheng, "Impact of high-order-mode loss on high-power fiber amplifiers," J. Opt. Soc. Am. B **33**(6), 1030–1037 (2016).
39. G. Fu, D. Li, M. Gong, P. Yan, and Q. Xiao, "Spatiotemporal deterioration in nonlinear ultrafast fiber amplifiers", Appl. Phys. Lett. **123**(9), 091106 (2023).
40. See Supplemental Material at ……for other simulation details, MMGF amplification, and numerical results of soliton attractors.